\documentclass[aps,prb,twocolumn,showpacs,groupedaddress]{revtex4}
\usepackage{graphicx}%

\begin{document}

\title{ Low frequency Raman studies of multi-wall carbon nanotubes:\\ experiments and theory }

\author{J. M. Benoit}
\altaffiliation{Present address: MPI-FK, Heisenbergstrasse 1, 70569 Stuttgart, Germany}
\author{J. P. Buisson}
\author{O. Chauvet}
\author{C. Godon}
\author{S. Lefrant }
\altaffiliation{ Corresponding author, e-mail: lefrant@cnrs-imn.fr }
\affiliation{Institut des Materiaux Jean Rouxel IMN/LPC, 2 rue de la Houssiniere, 44322 NANTES, FRANCE }

\date{\today}

\begin{abstract}
In this paper, we investigate the low frequency Raman spectra of multi-wall carbon nanotubes (MWNT) prepared by the electric arc method. Low frequency Raman modes are unambiguously identified on purified samples thanks to the small internal diameter of the MWNT. We propose a model to describe these modes. They originate from the radial breathing vibrations of the individual walls coupled through the Van der Waals interaction between adjacent concentric walls. The intensity of the modes is described in the framework of bond polarization theory. Using this model and the structural characteristics of the nanotubes obtained from transmission electron microscopy allows to simulate the experimental low frequency Raman spectra with an excellent agreement. It suggests that Raman spectroscopy can be as useful regarding the characterization of MWNT as it is in the case of single-wall nanotubes. 
\end{abstract}


\maketitle
Since their discovery, carbon nanotubes attract a huge interest in the scientific community due to their potential use in future devices, exploiting either their mechanical or electrical properties. Much attention is paid to single-wall nanotubes (SWNT) which exhibit electronic properties which depend on their type and/or chirality. Among the techniques extensively used to characterize these materials, Raman scattering plays a fundamental role as reviewed by Dresselhaus and Eklund in ref.\onlinecite{adv-phys}. On one hand, the radial-breathing mode (RBM) in the low frequency region shows a straightforward dependence on the diameter of the SWNT to a determination of the distribution of tube diameters\cite{nature}. On the other hand, strong resonance effects lead to different profiles of the tangential modes (G band), in particular in the red excitation range for which metallic tubes contribute significantly to the Raman response\cite{metal-SWNT}. Multi-wall nanotubes (MWNT) are nowadays also studied extensively and characterized by Raman scattering. Low frequency Raman modes have also been observed as well and resonance effects have been detected\cite{raman-MWNT,resonance}. However, up to now, the results have been rather dispersed and no clear interpretation of the experimental results has been offered.

In this paper, we present a detailed low frequency Raman study of arc discharge multi-wall carbon nanotubes which are characterized by transmission electron microscopy (TEM). We clearly identify low frequency modes in purified samples. We present a model to interpret these modes based on the coupling of the RBM of each individual wall of the MWNT. Intensity calculations are proposed as well. The low frequency spectrum of the MWNT is shown to be very sensitive to the internal diameter of MWNT. For the first time, a comprehensive simulation of low frequency Raman spectrum of MWNT is given. Preliminary results have already been presented elsewhere\cite{JP1}.

\begin{figure}
\includegraphics[bb=13 655 236 826, clip]{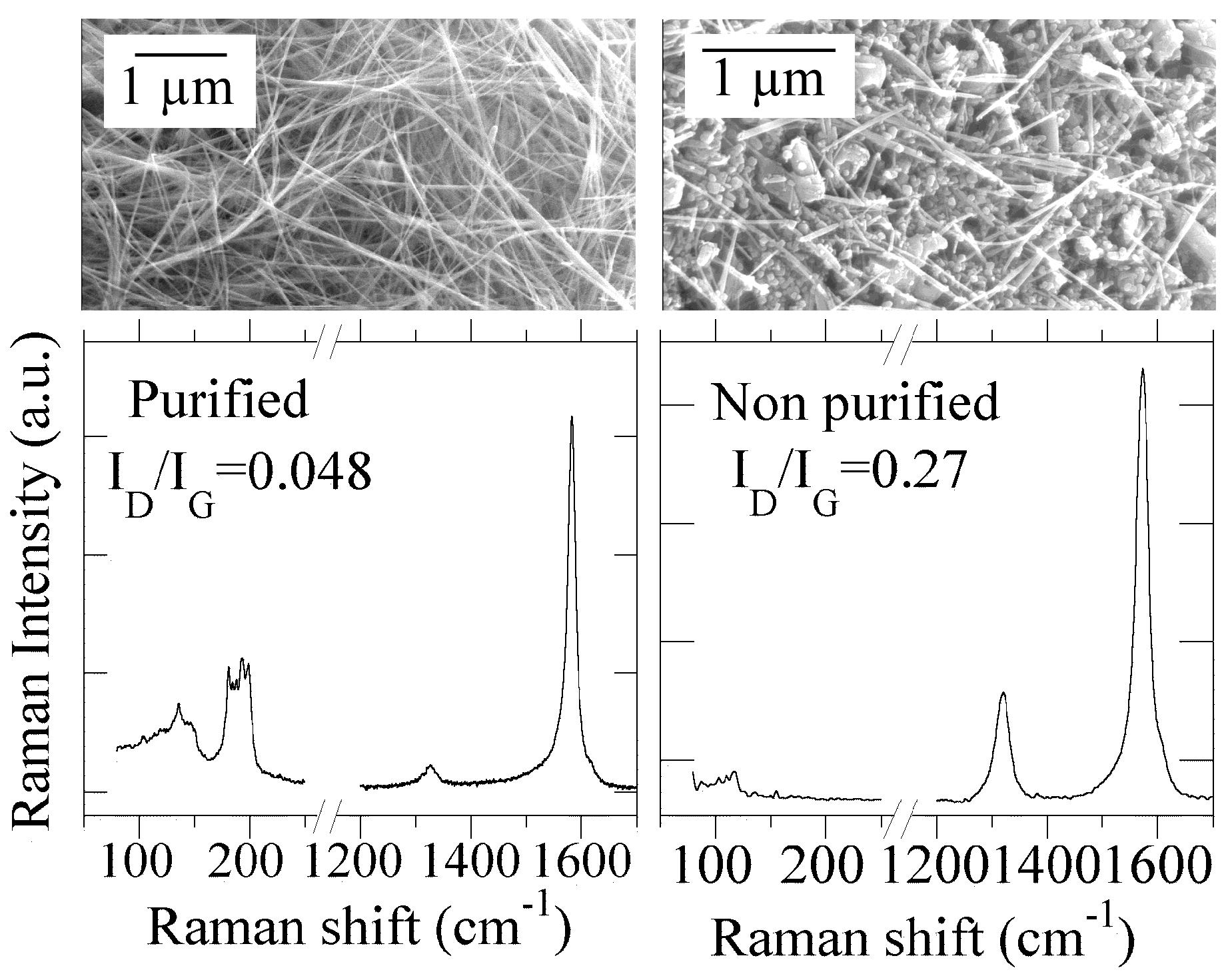}
\caption{\label{fig1} Upper part: SEM images of the MWNT material before and after the purification process. Lower part: Raman spectra ($\lambda_{exc}$= 676 nm) of the two MWNT materials. An important low frequency contribution is observed in the purified sample.}
\end{figure}
\begin{figure}
\includegraphics[bb=13 678 245 827, clip]{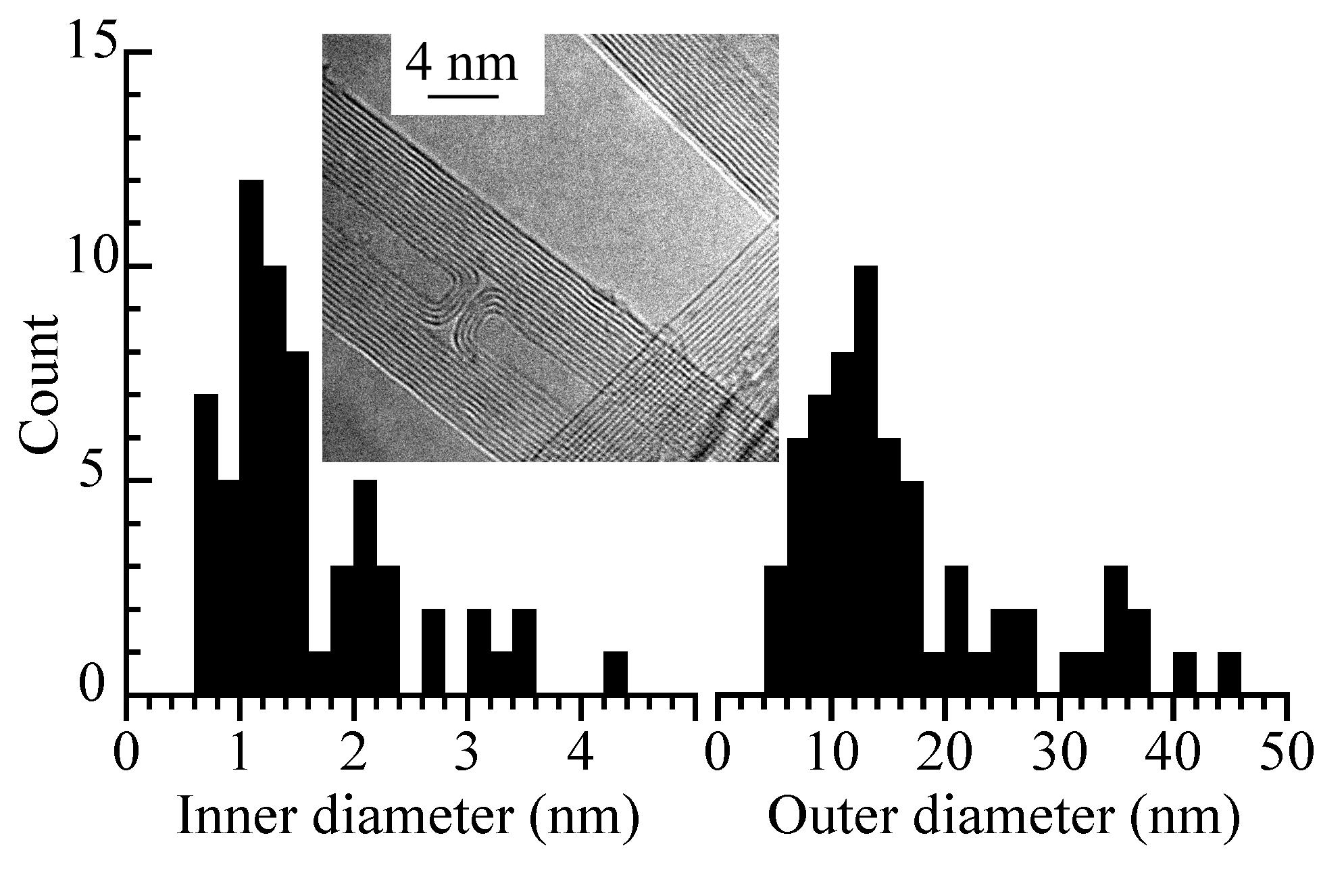}
\caption{\label{fig2} Histograms of the inner and outer diameter distribution of the MWNT, deduced from HRTEM observations. Inset: HRTEM image of the material which shows that the MWNT are highly graphitized.}
\end{figure}

The MWNT samples are synthesized by sublimation of pure graphite rods using electric arc discharge at Trinity College, Dublin using standard conditions. A typical micrograph of the material extracted from the cathode of the arc set-up and obtained by scanning electron microscopy (SEM) is given in the right part of Fig.\ref{fig1}. Carbon fibrils which constitute the MWNT as well as a rather large number of polyhedric particles are found. From SEM observations, the purity of the sample does not exceed 50 \%. In order to improve the quality of the samples, we follow the oxidation purification procedure described in ref.\onlinecite{purif}. As shown in Fig.\ref{fig1}, the sample quality is drastically improved even if some parts remains of poorer quality. High resolution TEM (Hitachi H9000 NAR) of the purified material confirms the SEM observations. It shows also that the MWNT are very well graphitized as shown in inset of Fig.\ref{fig2}. Indeed degradations and tube opening sometimes occur.
Fig.\ref{fig2} shows histograms of the diameter distributions obtained from TEM observations. The inner diameter distribution is peaked around 1.2 nm whereas the outer diameter one has a maximum close to 14 nm. On average, these MWNT have 18 walls.

Room temperature Raman spectra of our samples were obtained using a multichannel Jobin-Yvon T64000 spectrometer. A systematic observation of the investigated area is performed before any measurement using the CCD camera of the Raman microscope in order to probe the quality of the area. Typical Raman spectra using the $\lambda$= 676.4 nm excitation line are shown in Fig.\ref{fig1}. The two spectra correspond to the purified and the non purified SEM micrographs shown in the upper part of the figure. Looking to other areas with similar morphology gives similar results. The three different regions of interest are: the G band region close to 1600 cm$^{-1}$, the D band one close to 1330 cm$^{-1}$, and the low frequency range below 200 cm$^{-1}$ on which we focus on this paper. As noticed above, low frequency Raman modes have already been observed in MWNT\cite{raman-MWNT, resonance} although a large dispersion of results is observed. Fig.\ref{fig1} gives clues to understand this behavior. The low frequency modes are clearly observed in the high quality sample while they are not in the raw sample. This difference in sample quality is also reflected in the two other Raman contributions: in the purified sample, the D band is weaker and the G band narrower. A direct measurement of the D band to G band intensity ratio gives $I_{D}/I_{G}$ =0.048 after purification to be compared to $I_{D}/I_{G}$ = 0.27 before. Another quantity of interest is the full width at half maximum (FWHM) of the G band. Its value is 17 cm$^{-1}$ for the purified sample and 27 cm$^{-1}$ for the non purified one to be compared to 13 cm$^{-1}$ for HOPG graphite\cite{width}. In pre-graphitic carbons, both quantities, $I_{D}/I_{G}$ and FWHM, are known to depend on the size of the graphitic crystallites\cite{Tuinstra, Nakamura}: the smaller these parameters, the highest the degree of crystallinity. Here indeed, the reduction by one order of magnitude of $I_{D}/I_{G}$ and by almost a factor of 2 of the width of the G band in the purified sample has to be related with the destruction of most of the polyhedric particles as observed by SEM and presumably of a part of the amorphous carbon. Conversely, this work suggests that low frequency modes are not observed in the Raman spectra of most of the as-grown MWNT samples because they are hidden by the prominent contribution of other carbonaceous compounds. 

\begin{figure}
\includegraphics[bb=20 577 244 822, clip]{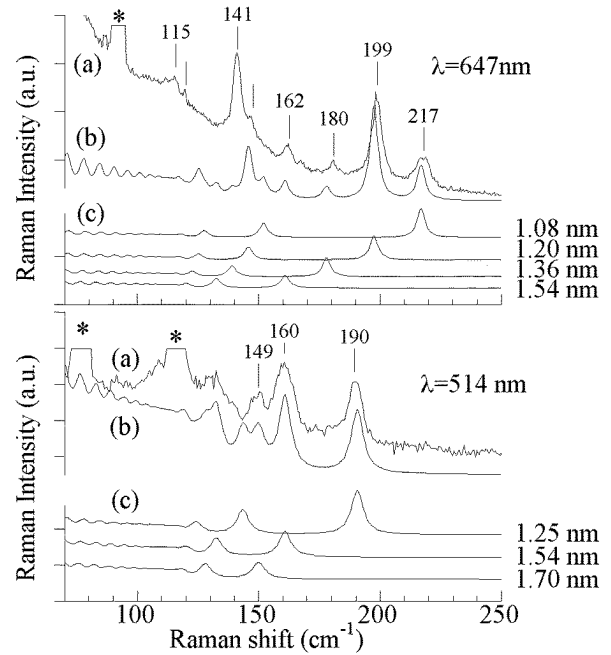}
\caption{\label{fig3} (a) Experimental Raman spectra of MWNT using $\lambda$=676 nm and $\lambda$=514 nm laser excitations on the same sample area. (b) Results of the simulation discussed in the text. We used a FWHM of 4 cm$^{-1}$ and 6 cm$^{-1}$ for the $\lambda$=676 nm and $\lambda$=514 nm spectra respectively. (c) Individual contributions of the 20 walls nanotubes involved in the simulation shown in (b). The inner diameter of the 20 walls tubes are given on the right. The star markers indicate laser lines. We do not substract any baseline on the experimental spectra.}
\end{figure}
Definitive assignment of the low frequency Raman modes to MWNT is confirmed by using a thin film of 30 \% purified MWNT mixed with polymethylmetacrylate. Observations with the Raman CCD camera and with SEM as well clearly show MWNT emerging from the fracture zones of the film while polyhedric particles are not observed in the same conditions. Raman studies at different excitation wavelengths can thus be performed on the same particle free area by focusing the laser beam on such a zone. Two low frequency spectra obtained using $\lambda$= 676 nm and and $\lambda$= 514 nm are presented in Fig.\ref{fig3}(a). Several low frequency modes are clearly visible from 115 to 220 cm$^{-1}$. The position of the peaks depends on the laser excitation suggesting resonance effects as pointed out in ref.\onlinecite{resonance}.

In SWNT, the low frequency part of the Raman spectrum is due to the radial breathing modes (RBM) with A$_{1g}$ symmetry. In this paper, we suggest that the low frequency modes observed in MWNT have the same origin, \textit{i.e.} they also originate from the breathing vibrations of the individual walls of the MWNT. However, because the individual walls are concentric, the situation is much more complex than in SWNT. Three ultimate assumptions can be made. \textit{i-} The different walls can "breathe" independently. It is very unlikely. As discussed below, despite its weakness, the Van der Waals interaction is sufficient to allow the coupling of the individual vibrations. Furthermore, we should expect to observe many more peaks than those observed in Fig.\ref{fig3}. \textit{ii-} Only the most outer wall is able to breathe. If so, we should not expect to observe modes in the 200 cm$^{-1}$ region but instead below 50 cm$^{-1}$. \textit{iii-} The different walls can breathe however the breathing vibrations are affected by the interactions between concentric walls. Here we propose a model to describe these interactions.

The problem of interacting RBM has already been considered theoretically to describe the "bundle effect" in SWNT\cite{bundle-effect}. In our group, we have addressed the "bundle effect" problem for an hexagonal lattice of SWNT\cite{JP1}. A Lennard-Jones potential is used to describe the carbon-carbon long range interaction. The tube-tube interaction is obtained from the mutual integration of the potential over adjacent SWNT. The second derivatives of the interaction give the tube-tube force constants. The adjacent tube-tube force constant $C$ is given by: $C =0.3 C_{g}\times(d_{VdW}/d)^{1/2}$ in which $d$ is the SWNT diameter, $d_{VdW}$ is the distance between adjacent SWNT and $C_{g}$=2.3 N/m corresponds to the force constant between two graphene sheets extracted from the B$_{1g}$ mode of graphite. In this simplified model, we consider a bundle of $N$ identical tubes. The $N$ individual RBM are coupled through the tube-tube interactions, resulting in $N$ new modes. The frequency of these modes is obtained by diagonalizing the dynamical matrix. This model gives an upshift of the RBM mode of (10,10) SWNT in bundle in the range 11-16 cm$^{-1}$, in agreement with other authors\cite{bundle-effect}.

Here we adapt this approach to the case of the low frequency Raman modes of a MWNT. A multi-wall nanotube is described by its inner diameter $d_{1}$ and its outer diameter $d_{N}$, $N$ being the number of walls. Each wall has a diameter $d_{i}$ and it can be characterized individually by its radial breathing mode $\omega_{0i}(cm^{-1})$=223.7/$d_{i}$ (nm)\cite{omega0-swnt}. In the following, the wall-wall distance is fixed at 0.34 nm. We estimate the concentric wall-wall interactions in the same way as before. We limit these interactions to first neighboring walls since extending the interactions to all the walls does not change significantly the final result. The first neighbor wall-wall interactions turn out to be independent of the wall diameters with force constants close to $C_{g}$ when the wall-wall distance is 0.34 nm. Introducing these interactions couples the $N$ individual modes, resulting in $N$ new modes. Again, the frequencies $\omega_{i}$ of the $N$ new modes are obtained after diagonalization of the dynamical matrix. As expected, all the new modes are upshifted by comparison with the bare modes. The upshift $\Delta_{i}=\omega_{i}-\omega_{0i}$ is the strongest for intermediate modes. For example, considering a $N$=20 MWNT with inner diameter $d_{1}$=1.2 nm, the most inner or outer modes are only shifted by $\Delta_{1}$=+12 cm$^{-1}$ and $\Delta_{N}$=+3 cm$^{-1}$, respectively, while a shift of $\Delta_{i}$=+70 cm$^{-1}$ is obtained for modes with $i$ close to 10. Still, this model is not sufficient to explain the experimental data, and more precisely the fact that we observe well defined and intense modes in the 200 cm$^{-1}$ region. Actually, the dynamical matrix diagonalization shows that in-phase vibration is achieved for the mode number $N$, \textit{i.e.} for the mode at the lowest frequency. It suggests that the Raman intensity of the modes should be small at the highest frequencies. 

In order to address this problem carefully, we propose also an intensity calculations. We do not take into account any possible resonance effect. The polarisability tensor of each individual wall is obtained using the non-resonant bond polarization theory\cite{polarization}. It turns out that the derivative polarisability tensor d($\alpha$)/d($d_{i}$) of an isolated wall with diameter $d_{i}$ scales with ${d_{i}}^{-1/2}$, then giving rise to a $\omega_{0i}$ RBM of intensity proportional to 1/$d_{i}$. In the whole MWNT, the derivative polarisability tensor is equal to the sum of the tensors of the individual wall multiplied by the amplitude of the vibration of the considered mode. The results of this calculation show that the most intense modes are those which originate from the vibration of the most internal and external walls, respectively. In fact, this is due to two competitive effects. \textit{i-} The smallest walls contribute to the intensity as 1/$d_{i}$. \textit{ii-} The smallest frequency mode (originating from the biggest wall) is associated with an in-phase vibration, inducing there a cumulative effect in term of intensity. Such calculations reveal that two parameters are of primary importance, the internal tube diameter on one hand, and the number of walls in the MWNT on the other hand. Such data can be statistically extracted from TEM observations.

\begin{figure}
\includegraphics[bb=20 675 226 823, clip]{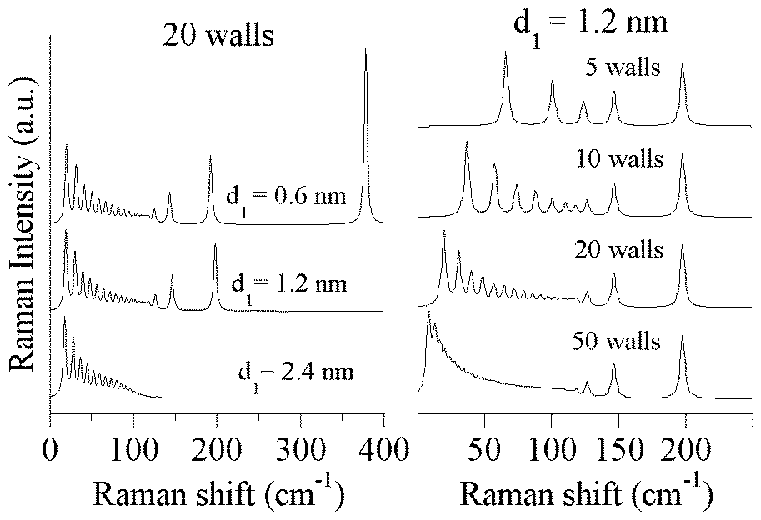}
\caption{\label{fig4} Simulated low frequency Raman spectra of MWNT. The FWHM of each peak is arbitrary fixed to 4 cm$^{-1}$. Left pannel: effect of the inner diameter $d_{1}$ on a 20-walls tube. Right pannel: effect of the number of walls on a $d_{1}$=1.2 nm MWNT whose number of walls varies between 5 to 50. }
\end{figure}
The combine results of both the frequency and intensity calculations are shown in Fig.\ref{fig4} where we arbitrary use a FWHM=4 cm$^{-1}$. As shown in the left panel of the figure, the modes at highest frequencies originate from the smallest wall. However, their intensities become vanishingly small as soon as the inner diameter exceeds 2 nm. It suggests that low frequency modes of MWNT with large inner diameter cannot be observed even if present. It may explain partly the disparity of the results published in the literature. In the right panel of the figure, we simulate the spectrum of a MWNT with inner diameter $d_{1}$=1.2 nm and $N$ walls. The intensity evolution described before is clearly visible. More interestingly, we can divide the spectra in two parts. The three modes at highest frequencies are only weakly affected by the number of walls. However, when looking to lower frequencies ($\omega_{i}\leq$130 cm$^{-1}$), the situation is dramatically different when $N$ increases from 5 to 50. This is due to two facts. First, the number of modes in the 0-130 cm$^{-1}$ is increasing with $N$, giving rise to the oscillatory behavior. Then, the upshift $\Delta_{i}$ of the intermediate modes is also increasing with $N$ pushing all the modes closer to each other.

This model can be used to simulate the experimental spectra of Fig.\ref{fig3}. We recall that our calculations do not take into account any possible resonance effects. From the TEM results, we consider MWNT with inner diameters between 1 and 2 nm and with $N$=20. Let us start with the spectrum at $\lambda$= 647 nm. Four well resolved peaks are observed at 217, 199, 180 and 162 cm$^{-1}$ respectively. Two doublets are observed close to 141 and 115 cm$^{-1}$ respectively. We need four MWNT with inner diameter 1.08, 1.20, 1.36 and 1.54 nm respectively to yield the four modes at highest frequencies. The calculated spectra of this four MWNT are given in part (c) of the figure with the same intensity unit. The resulting simulated spectrum is shown as spectrum (b) in the figure and using the relative weights 16 \%, 55 \%, 10 \% and 19 \% for the MWNT with $d_{1}$=1.08, 1.20, 1.36 and 1.54 nm respectively (the relative weights take into account the intensity considerations discussed before). Indeed, the simulated spectrum describes very well the highest frequency modes. Much more interestingly, the simulated spectrum exhibits two doublets, one close to 145 cm$^{-1}$, the second one close to 125 cm$^{-1}$ whose intensity are comparable to the two experimental doublets. The origin of these doublets is easily found in the individual spectra shown in part (c). The first one comes from the second modes of the $d_{1}$=1.08 and $d_{1}$=1.20 nm MWNT and the second one from the third modes. Indeed, this simulation strongly supports our model. Still, a small discrepancy in the position of the doublets remains, this discrepancy being bigger when the third modes are involved. Since the experimental doublets are at lower frequencies, it may indicate that the involved MWNT have less than 20 walls. The second experimental spectrum was obtained on the same sample area with $\lambda$= 514 nm. Two peaks are found at 190 and 160 cm$^{-1}$ and unresolved features appear close to 149 and 130 cm$^{-1}$. The simulated spectrum shown in (b) is obtained by considering MWNT with $d_{1}$=1.25, 1.54 and 1.70 nm and relative weights 22 \%, 45 \% and 33 \% respectively. In this case, the simulated spectrum is able to describe the experimental unresolved feature at 149 cm$^{-1}$ and the second one at 130 cm$^{-1}$, due to the second mode contributions. There is less discrepancy in frequency positions than in the previous case. This is what is expected if the number of walls is less than the one we used.

In summary, low frequency Raman modes in MWNT have been unambiguously identified provided the MWNT are of high quality and their internal diameter is small (less than 2 nm). These modes which originate from the radial breathing modes of the individual walls are strongly coupled through the concentric tube-tube Van der Waals interaction. Frequency and intensity calculations performed in the framework of the dynamical matrix diagonalization and non resonant bond polarization theory allow to simulate experimental Raman spectra with a rather good agreement. Still, resonance effects are observed which are not taken into account in our model. They deserve further experimental and theoretical work.\\This work is partly supported by the EEC COMELCAN HPRN-CT-2000-00128 contract

\end{document}